\documentclass[12pt,a4paper,final]{iopart}

\usepackage{iopams}

\usepackage{graphicx}
\usepackage{dcolumn}
\usepackage{float}
\usepackage{iopams}
\usepackage{multirow}
\usepackage[breaklinks=true,colorlinks=true,linkcolor=blue,urlcolor=blue,citecolor=blue]{hyperref}

\begin{document}

\title{Electron-boson spectral density function of correlated multiband systems obtained from optical data: Ba$_{0.6}$K$_{0.4}$Fe$_2$As$_2$ and LiFeAs}

%% Notice placement of commas and superscripts and use of &
%% in the author list
\author{Jungseek Hwang}

\address{$^1$Department of Physics, Sungkyunkwan University, Suwon, Gyeonggi-do
440-746, Republic of Korea}

\ead{jungseek@skku.edu}

\date{\today}

\begin{abstract}
We introduce an approximate method which can be used to simulate the optical conductivity data of correlated multiband systems for normal and superconducting cases by taking advantage of a reverse process of a usual optical data analysis, which has been used to extract the electron-boson spectral density function from measured optical spectra of single-band systems, like cuprates. We applied this method to optical conductivity data of two multiband pnictide systems (Ba$_{0.6}$K$_{0.4}$Fe$_2$As$_2$ and LiFeAs) and obtained the electron-boson spectral density functions. The obtained electron-boson spectral density consists of a sharp mode and a broad background. The obtained spectral density functions of the multiband systems show similar properties as those of cuprates in several aspects. We expect that our method helps to reveal the nature of strong correlations in the multiband pnictide superconductors.
\end{abstract}


\maketitle

%
%\bigskip

\section{Introduction}

Inelastic charge carrier scattering spectra of superconducting materials may contain crucial information on the force of forming the Cooper pairs. Particularly for conventional superconductors this quantity has been successful to expose the electron-phonon density (or glue) spectrum\cite{carbotte:1990}. Optical spectroscopy has played an important role to reveal the electron-phonon density spectrum along with tunneling and inelastic neutron scattering spectroscopies\cite{mcmillan:1965,stedman:1967,farnworth:1974}. For high-temperature superconducting cuprate systems the electron-boson spectral density has been obtained by various experimental techniques\cite{carbotte:2011}. The mediated bosons in cuprates have been associated with the antiferromagnetic spin-fluctuations and there are a lot of experimental evidences and theoretical interpretations to support the idea\cite{dai:1999,carbotte:1999,hwang:2004,maier:2008,dahm:2009}.

Iron-pnictides discovered by Kamihara {\it et al.}\cite{kamihara:2006} have been known to have multiband nature\cite{subedi:2008,ding:2008,fang:2009,wu:2010}. For multigap superconducting (or multiband) systems the process for extracting the electron-boson spectral density functions will be more complex compared with that for cuprates (one-band systems) which have a single $d$-wave superconducting gap. Many optical studies have been successfully done on cuprate systems for delving and exposing the electron-boson spectral density function\cite{carbotte:1999,schachinger:2000,dordevic:2005,hwang:2006,hwang:2007,schachinger:2008,heumen:2009}. Recently some optical studies\cite{yang:2009a,wu:2010b,hwang:2015} have been done on iron-pnictides to obtain the electron-boson density spectrum by using a single-band approximation method, which has been used for studying cuprates. Charnukha {\it et al.}\cite{charnukha:2011} have tried to include the multiband coupling information obtained from specific heat measurement study of the same material\cite{popovich:2010} and showed that observed unusual structure of the optical conductivity in superconducting state can be ascribed to spin-fluctuation-assisted process. However, their method cannot be applicable to analyze optical spectra directly. As far as the author knows reliable methods for obtaining the electron-boson spectral density from optical data of multiband systems, iron-pnictides, have not been developed yet.

In this paper we propose an approximate method which can be used to obtain the electron-boson spectral density function from optical data of correlated multiband superconductors by taking advantage of a reverse process reported recently by Hwang\cite{hwang:2015a}. We denote the electron-boson spectral density function as $I^2\chi(\omega)$, where $I$ is the coupling constant between an electron and a mediating boson and $\chi(\omega)$ is the mediating boson spectrum. We started from two typical model $I^2\chi(\omega)$ functions, calculated the imaginary parts of the optical self-energy using the Allen's formulas for both normal and superconducting cases (here we need to know the impurity scattering rate and additionally the superconducting gap for superconducting case), obtained the corresponding real parts of the optical self-energy using a Kramers-Kronig relation between imaginary and real parts of the optical self-energy, calculated the optical conductivity spectra using an extended Drude model (here we need to know the plasma frequency) from the calculated optical self-energy. We can get other optical constants from the optical conductivity using the relations between them. We repeated the whole process mentioned above for another transport channel with the same input $I^2\chi(\omega)$, different impurity scattering rate, plasma frequency, and superconducting gap. Here we made a rather radical assumption that $I^2\chi(\omega)$ is the same for each channel; we speculated that bosonic fluctuations which contribute to the two channels are the same and play the major role for the correlation between charge carriers in both channels. We note that the channels are closely related to two different superconducting gap channels. Then eventually we could obtain a combined optical conductivity of the two charge-transport channels by adding the two optical conductivity data obtained using the reverse process. Then we compared the resulting combined optical conductivity with the experimental measured optical data and, interestingly, found that the combined conductivity could capture some characteristic features in the experimental data. Therefore, we applied this method to the measured optical conductivity data through a fitting process and obtained the electron-boson spectral density functions of two multiband pnictide systems: one is an optimally doped Ba$_{0.6}$K$_{0.4}$Fe$_2$As$_2$ and the other LiFeAs. We compared the obtained electron-boson spectral density functions obtained from pnictides optical data with those of cuprates and found that the spectral density functions of two different material systems showed common properties in several aspects.

\section{Formalism}

\subsection{The reverse process and combined optical conductivity of two channels}
In a (multiband) correlated material system the $k$-space averaged electron-electron interaction through mediated bosons can be specified by an electron-boson spectral density function ($I^2\chi(\omega)$), which can be measured by various spectroscopic techniques including optical spectroscopy. In this section we introduce a reverse process\cite{hwang:2015a}, which we use in this paper. In this process we start from an input $I^2\chi(\omega)$. And the first step is that we obtain the imaginary part of the optical self-energy ($\tilde{\Sigma}^{op}(\omega) \equiv \Sigma^{op}_1(\omega) + i\Sigma^{op}_2(\omega)$) from an input $I^2\chi(\omega)$ using a generalized Allen's formula\cite{allen:1971} as
\begin{equation}\label{eq1}
\Sigma^{op}_2(\omega,T)=-\frac{1}{2}\Big{[}\int^{\infty}_{0}\!\!d\Omega \:I^2\chi(\omega,T) \:K(\omega,\Omega,T) + \frac{1}{\tau^{op}_{imp}(\omega)}\Big{]},
\end{equation}
where $K(\omega,\Omega,T)$ is the kernel of the Allen's integral equation and $1/\tau^{op}_{imp}(\omega)$ is the impurity scattering rate, which is a constant for normal case and is frequency-dependent near the superconducting gap for superconducting case\cite{allen:1971,hwang:2015a}. In our simulations we used the kernels of normal ($T =$ 0 K) and $s$-wave superconducting states\cite{hwang:2015a} as
\begin{eqnarray}\label{eq2}
K(\omega,\Omega) &=& \frac{2\pi}{\omega}(\omega-\Omega)\Theta(\omega-\Omega) \:\:\:(\mbox{for normal state}) \nonumber \\ &=& \frac{2\pi}{\omega}(\omega-\Omega)\Theta(\omega-2\Delta_0 - \Omega) \nonumber \\ \!\!&\times& \!\!\! E\Big{(}\frac{\sqrt{(\omega-\Omega)^2\!-\!(2\Delta_0)^2}}{\omega\!-\!\Omega}\Big{)} (\mbox{for $s$-wave SC}),
\end{eqnarray}
where $\Theta(\omega)$ represents the Heaviside step function, i.e., 1 for $\omega \geq 0$ and 0 for $\omega < 0$), $E(z)$ represents the complete elliptic integral of the second kind, where $z$ is dimensionless. Here we note that for superconducting case the formula was derived for $T =$ 0 K and for a finite temperature ($0 < T < T_c$) one can use the formula approximately with the superconducting gap at that temperature instead of $\Delta_0$, which is the maximum SC gap at $T =$ 0 K. The impurity scattering rates [$1/\tau^{op}_{imp}(\omega)$]\cite{allen:1971} for normal and $s$-wave superconducting states can be written as
\begin{eqnarray}\label{eq3}
\frac{1}{\tau^{op}_{imp}(\omega)} &=& \frac{1}{\tau_{imp}} \:\:\:(\mbox{for normal state}) \nonumber \\ \!\!&=&\!\! \frac{1}{\tau_{imp}} E\Big{(}\frac{\sqrt{\omega^2\!-\!(2\Delta_0)^2}}{\omega}\Big{)} \:(\mbox{for $s$-wave SC}),
\end{eqnarray}
where $1/\tau_{imp}$ is a constant impurity scattering rate. In our data fitting we used the so-called Shulga's kernel ($K(\omega, \Omega, T)$)\cite{shulga:1991}, which is valid for finite temperature ($T >$ 0) with a constant density of states in normal state: $K(\omega, \Omega, T) = 2\omega \coth(\Omega/2T)-(\omega+\Omega)\coth[(\omega+\Omega)/2T]+(\omega-\Omega)\coth[(\omega-\Omega)/2T]$, where $T$ is the temperature. We note that $K(\omega, \Omega, 0)$ is the same as the Allen's kernel for normal state in Eq. (\ref{eq2}). And then we obtained the real part of the optical self-energy ($\Sigma^{op}_1(\omega)$) from the calculated imaginary part ($\Sigma^{op}_2(\omega)$) using a Kramers-Kronig relation as
\begin{equation}\label{eq4}
\Sigma^{op}_1(\omega) = -\frac{2\omega}{\pi} P\int^{\infty}_{0}\frac{\Sigma^{op}_2(\Omega)}{\Omega^2-\omega^2}d\Omega,
\end{equation}
where $P$ stands for the principle part of the improper integral. We note that for the Kramers-Kronig integration we produced data in a wide spectral range from zero to around 7.1 eV, which seems to be large enough for our purpose (here we consider spectra up to 0.3 eV for model calculations and up to 0.1 eV for applications to real systems). For the higher energy up to infinity we extended the data with a gradual drop up to $\sim$124 eV with an exponent (-1) of frequency and after then a free electron behavior. Finally we get the complex optical conductivity ($\tilde{\sigma}(\omega) = \sigma_1(\omega) + i\sigma_2(\omega)$) using an extended Drude model as
\begin{equation}\label{eq5}
\tilde{\sigma}(\omega)=\frac{\omega_p^2}{8\pi i}\frac{1}{\tilde{\Sigma}^{op}(\omega)-\omega/2},
\end{equation}
where $\omega_p$ is the plasma frequency of charge carriers, which can be specified for each transport channel, and $\tilde{\Sigma}^{op}(\omega)$ $(=  \Sigma^{op}_1(\omega) + i\Sigma^{op}_2(\omega))$ is the complex optical self-energy. In principle, using this complex optical conductivity we can calculate any other optical constants including reflectance spectra\cite{hwang:2015a}.

Now we perform the same reverse process to get another complex optical conductivity for another charge transport channel with the same $I^2\chi(\omega)$ and different parameters: impurity scattering rate and plasma frequency for normal case, and additionally superconducting gap for superconducting case. Then we get the combined optical conductivity of two transport channels, which can be related to two separate bands in the Fermi surface, as
\begin{equation}\label{eq6}
\tilde{\sigma}_{total}(\omega) = \tilde{\sigma}_{ch1}(\omega) + \tilde{\sigma}_{ch2}(\omega),
\end{equation}
where $\tilde{\sigma}_{ch1}(\omega)$ and $\tilde{\sigma}_{ch2}(\omega)$ are the complex optical conductivities of the channel 1 and channel 2, respectively. We can simulate some characteristic structures in measured optical conductivity spectra by using an input electron-boson spectral density and appropriate fitting parameters for the two channels. Therefore, we applied this approximate method to optical data of multiband systems and obtained the electron-boson spectral density. The detailed description of the fitting process is given in the section of "Applications to real systems".

\subsection{The combined optical quantities from the fitting}
One can calculate other optical quantities from the combined optical conductivity obtained using our analysis (or fitting) method. The fitted combined complex optical conductivity can be written as $\tilde{\sigma}_{fit}(\omega) = \sigma_{fit,1}(\omega)+i\sigma_{fit,2}(\omega)$, where the real and imaginary parts can be obtained straightforward through the extended Drude formalism using the fitting parameters for each channel. The fitted complex dielectric function can be calculated using the general relationship between the optical conductivity and the dielectric function as
\begin{equation}\label{eq7}
\tilde{\epsilon}_{fit}(\omega)=\epsilon_H + i\frac{4\pi}{\omega}\tilde{\sigma}_{fit}(\omega),
\end{equation}
where $\epsilon_H$ is the background dielectric constant and $\tilde{\epsilon}_{fit}(\omega) \equiv \epsilon_{fit,1}(\omega) + i\epsilon_{fit,2}(\omega)$. The further calculations for reflectance ($R_{fit}(\omega)$) and the optical scattering rate ($1/\tau^{op}_{fit}(\omega)\equiv -2\Sigma^{op}_{fit,2}(\omega)$) can be done using the Fresnel's equation and the extended Drude formalism, respectively.

\section{Model calculations}

\begin{figure}[]
  \vspace*{-0.7 cm}%
  \centerline{\includegraphics[width=3.5 in]{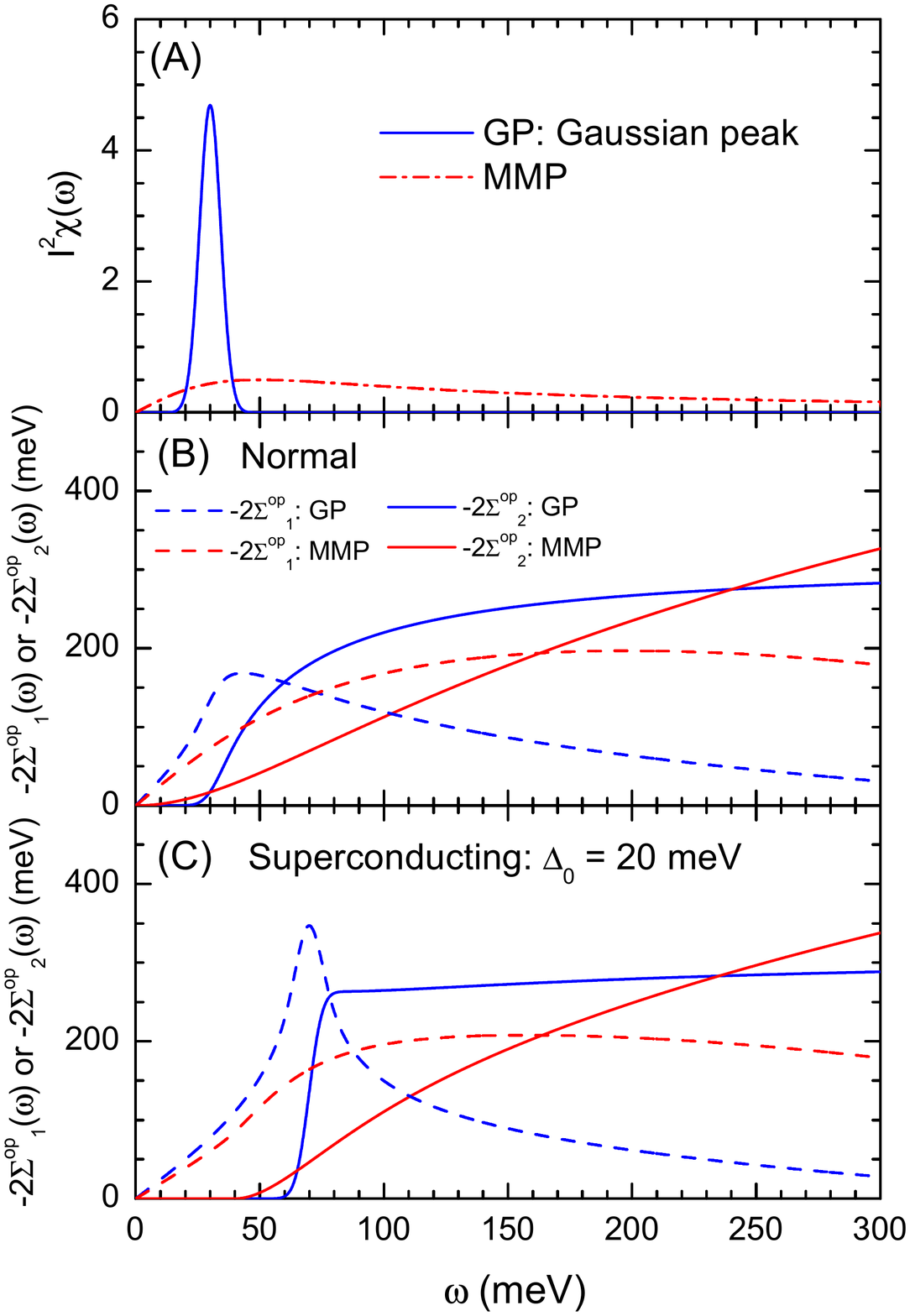}}%
  \vspace*{-1.0 cm}%
\caption{(A) Two model input electron-boson spectral density functions, $I^2\chi(\omega)$: a Gaussian peak (GP) and an MMP model. (B) The calculated real parts (dashed lines) and imaginary parts (solid lines) of the optical self-energy for the two model $I^2\chi(\omega)$ in normal case. (C) The calculated real parts (dashed lines) and imaginary parts (solid lines) of the optical self-energy for the two model $I^2\chi(\omega)$ in superconducting case.}
 \label{fig1}
\end{figure}

Now we introduce the way how we obtained the combined optical conductivity of two transport channels starting from an input model $I^2\chi(\omega)$ using the formalism introduced in the previous section. We used two input model $I^2\chi(\omega)$ functions shown in Fig. \ref{fig1}(A): one is a sharp Gaussian peak (GP) model (solid line: $A_{gp}/[\sqrt{2\pi} (d/2.35)] \exp{\{-(\omega-\omega_0)^2/[2 (d/2.35)^2]\}}$ peaked at $\omega_0$ (30 meV) with the area ($A_{gp}$) of 50 meV and the width ($d$) of 10 meV and the other the well-known MMP (Millis-Monien-Pines) model\cite{millis:1990} (dash-dotted line: $A_m\:\omega/(\omega^2+\omega_m^2)$) with a broad peak at $\omega_m$ (50 meV), the amplitude ($A_m$) of 50 meV and the cutoff frequency of 300 meV. We calculated the imaginary part of the optical self-energy from the input $I^2\chi(\omega)$ using the Allen's formula (Eq. (\ref{eq1}) and Eq. (\ref{eq2})). We note that here we did not include the impurity scattering rates since we will include them later on. The real parts are obtained using the Kramers-Kronig relation (Eq. (\ref{eq4})). The calculated real and imaginary parts of the optical self-energy of the two model $I^2\chi(\omega)$ functions for normal ($T =$ 0 K) and superconducting (with the superconducting gap, $\Delta_0 =$ 20 meV) cases are displayed in Fig. \ref{fig1}(B) and Fig. \ref{fig1}(C), respectively. The sharp Gaussian peak (GP) model causes pronounced peak features at $\omega_0$ (30 meV) and $\omega_0 + 2\Delta_0$ (70 meV) for normal and superconducting cases, respectively, and the corresponding imaginary parts show steep step-like features where the real parts show the peak features, which is consistent with reported results\cite{hwang:2008a,hwang:2015a}. The MMP model also shows similar but less pronounce features as those for the Gaussian peak model since the MMP model has a broad peak. In principle the saturation of the optical scattering rate in high energy region is $2 \pi$ times the area under $I^2\chi(\omega)$. The MMP model gives higher saturation level than GP models since the area ($\simeq$ 90 meV) under the MMP is larger than that (50 meV) of the GP model. We note that, in general, the optical scattering rate reaches its saturation at well above the cutoff frequency of $I^2\chi(\omega)$.

\begin{figure}[]
  \vspace*{-0.7 cm}%
  \centerline{\includegraphics[width=3.5 in]{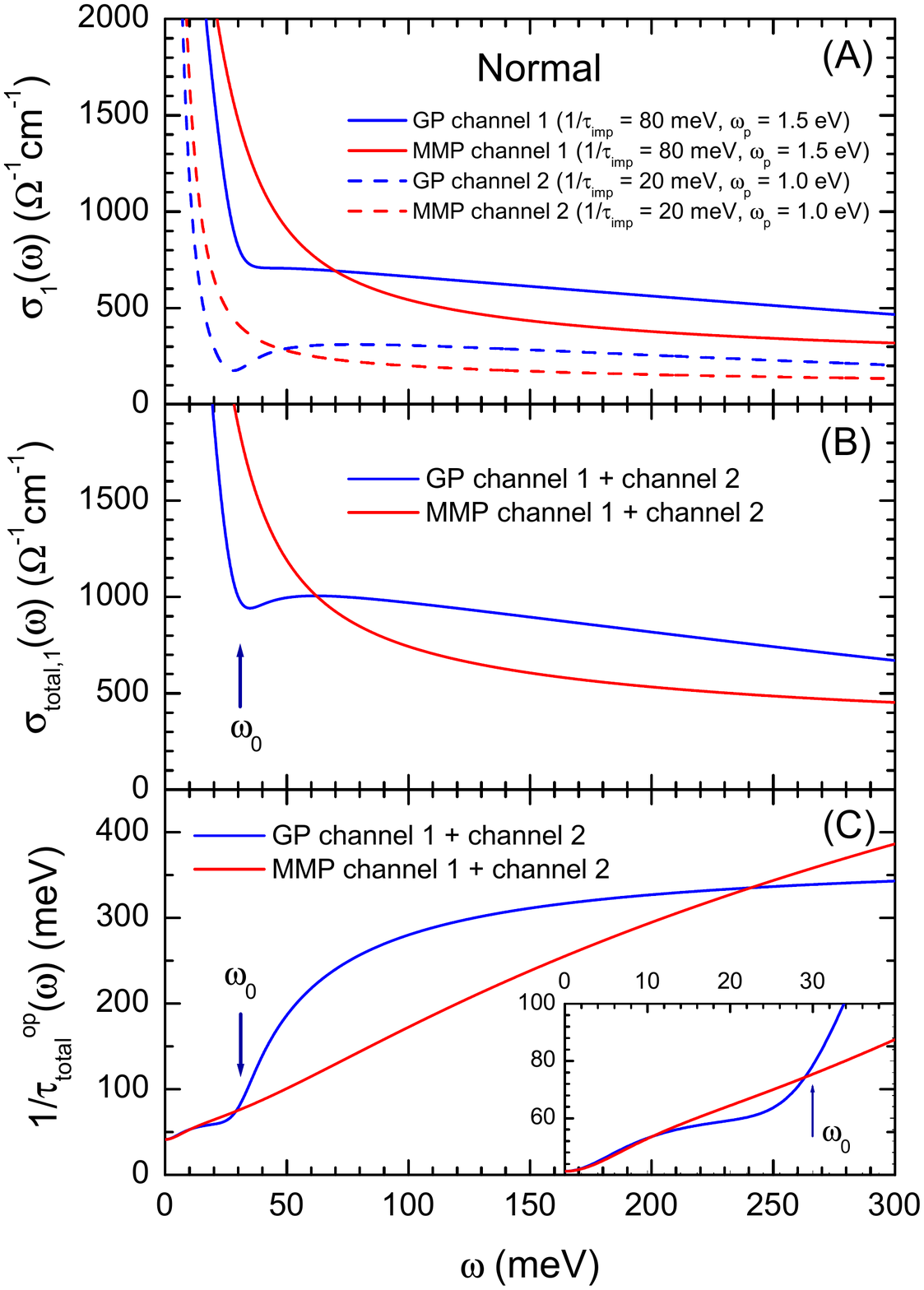}}%
  \vspace*{-1.0 cm}%
\caption{(A) The calculated optical conductivity spectra of two channels with different impurity scattering rates and plasma frequencies for normal case ($T =$ 0 K). (B) The combined optical conductivity spectra for the two model $I^2\chi(\omega)$ functions (GP and MMP). (C) The combined optical scattering rates for the two model $I^2\chi(\omega)$ functions. The arrows indicate the Gaussian peak frequency ($\omega_0$) of 30 meV. In the inset we show a magnified view in low frequency region below 40 meV.}
 \label{fig2}
\end{figure}

In Fig. \ref{fig2}(A) we display the real parts of the optical conductivity spectra of two transport channels ($\sigma_{ch1,1}(\omega)$: solid lines and $\sigma_{ch2,1}(\omega)$: dashed lines) for the two model $I^2\chi(\omega)$ functions at normal state ($T =$ 0 K). We take the plasma frequency, $\omega_{p,1} =$ 1.5 eV and the impurity scattering rate, $1/\tau_{imp,1} =$ 80 meV for the channel 1 and $\omega_{p,2} =$ 1.0 eV and $1/\tau_{imp,2} =$ 20 meV for the channel 2 to obtain the optical conductivity from the calculated complex optical self-energy using the extended Drude model (Eq. (\ref{eq5})). We note that here we included the impurity scattering rates in the imaginary part of the optical self-energy, i.e., $-2\Sigma^{op}_2(\omega) + 1/\tau^{op}_{imp}$, where $1/\tau^{op}_{imp}$ is a constant for normal case). In Fig. \ref{fig2}(B) we display the combined optical conductivity spectra ($\sigma_{total,1}(\omega) \equiv  \sigma_{ch1,1}(\omega) + \sigma_{ch2,1}(\omega)$) of the two channels for the two $I^2\chi(\omega)$ functions. The combined spectra do not seem to show any additional features which might be caused by the different two channels because the two channels show similar curve shapes and characteristic features at the same frequencies as shown in Fig. \ref{fig2}(A). However, the corresponding combined imaginary parts of the optical self-energy (or the optical scattering rates, $1/\tau_{total}^{op}(\omega) \equiv -2\Sigma^{op}_{total,2}(\omega)$) obtained using the extended Drude model show additional structures below $\sim$30 meV as shown in Fig. \ref{fig2}C and the inset of Fig. \ref{fig2}(C). Here we used the combined plasma frequency ($\omega_{p,total} \equiv \sqrt{\omega_{p,1}^2+\omega_{p,2}^2}$, $\simeq$ 1.8 eV) to obtain the combined scattering rate. We can understand the additional structure if we consider that while the two conductivities can be added linearly the corresponding impurity scattering rates cannot be done so. We note that the additional structure is not an issue of this paper. A characteristic feature appears at the sharp Gaussian peak frequency ($\omega_0$) marked with a magenta arrow in both combined conductivity and scattering rate spectra. In general this characteristic frequency divides the optical conductivity into two parts: the coherent and incoherent parts which are caused by the correlation effect\cite{hwang:2008a}.

\begin{figure}[]
  \vspace*{-0.7 cm}%
  \centerline{\includegraphics[width=3.5 in]{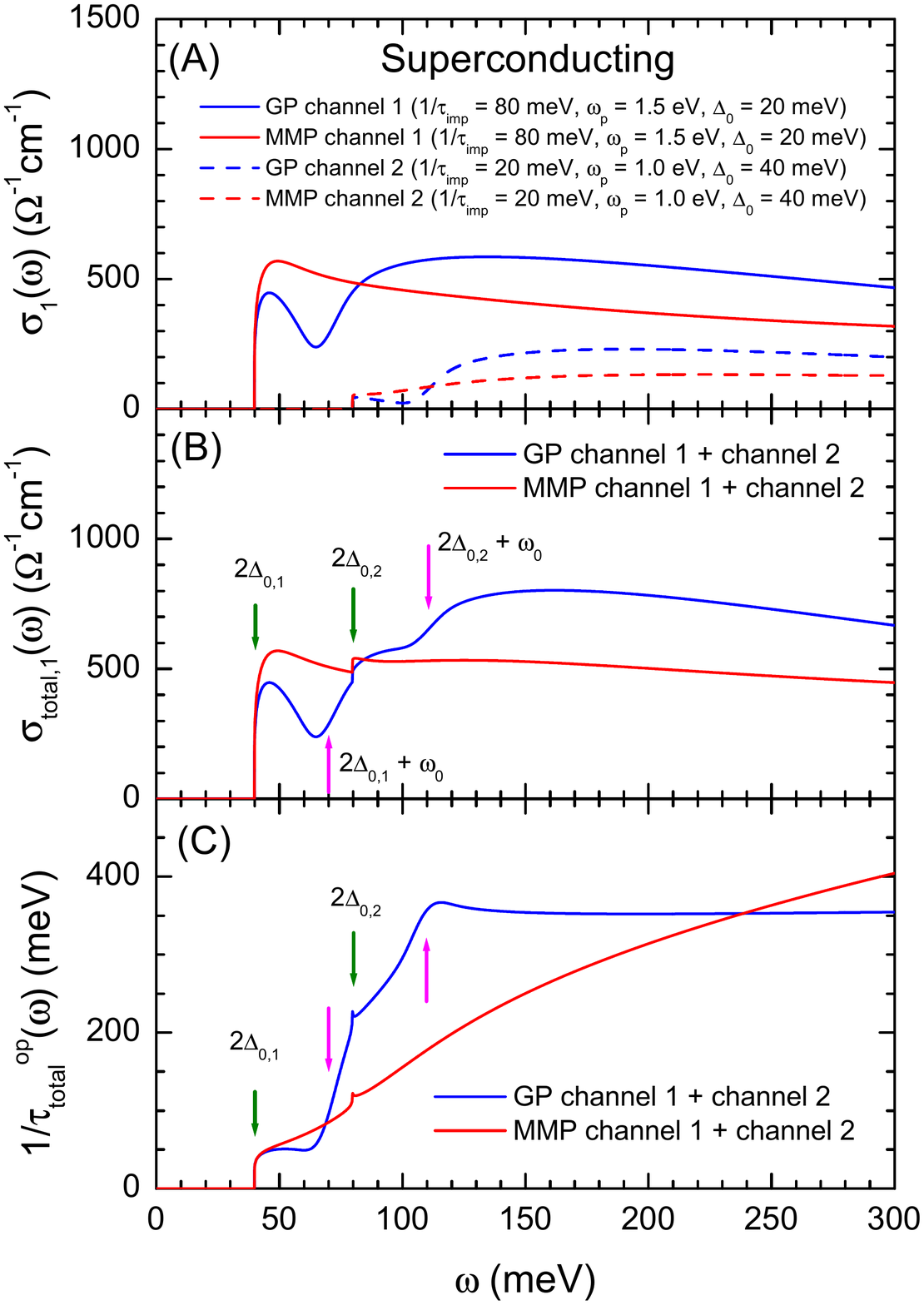}}%
  \vspace*{-1.0 cm}%
\caption{(A) The calculated optical conductivity spectra of the two channels with different impurity scattering rates, plasma frequencies, and the superconducting gaps for superconducting case. (B) The combined optical conductivity spectra for the two model $I^2\chi(\omega)$ functions. The arrows indicate characteristic energy scales: the superconducting gaps ($2\Delta_{0,i}$) and the Gaussian peak plus the SC gaps ($\omega_0+2\Delta_{0,i}$), where $i =$ 1 or 2. (C) The combined optical scattering rates for the two model $I^2\chi(\omega)$ functions.}
 \label{fig3}
\end{figure}

In Fig. \ref{fig3}(A) we display the optical conductivity spectra of two transport channels ($\sigma_{ch1,1}(\omega)$: solid lines and $\sigma_{ch2,1}(\omega)$: dashed lines) for the two model $I^2\chi(\omega)$ functions at superconducting state. We take $\omega_{p,1} =$ 1.5 eV, $1/\tau_{imp,1} =$ 80 meV and $\Delta_{0,1} =$ 20 meV (SC gap) for the channel 1 and $\omega_{p,2} =$ 1.0 eV, $1/\tau_{imp,2} =$ 20 meV and $\Delta_{0,2} =$ 40 meV (SC gap) for the channel 2 to obtain the optical conductivity from the calculated complex optical self-energy using the extended Drude model (Eq. (\ref{eq5})). We note that here we included the impurity scattering rates in the imaginary part of the optical self-energy for each channel, i.e., $-2\Sigma^{op}_2(\omega) + 1/\tau^{op}_{imp}(\omega)$, where $1/\tau^{op}_{imp}(\omega)$ is frequency-dependent for SC case as in Eq. (\ref{eq3})). We can see clearly some pronounced features caused by the superconducting gaps at $2\Delta_{0,i}$ and by both the Gaussian peak and the SC gaps at $\omega_0 + 2\Delta_{0,i}$, where $i =$ 1 or 2. In Fig. \ref{fig3}(B) we display the combined optical conductivity data for the two model $I^2\chi(\omega)$ functions. We can see the two channel contributions separately in the combined conductivity since they have different superconducting gaps; the two superconducting gaps give different onset frequencies ($2\Delta_{0,1}=$ 40 meV for channel 1 and $2\Delta_{0,2} =$ 80 meV for channel 2) as we marked with dark green arrows. We also can clearly see features caused by the sharp Gaussian peak combined with the SC gaps ($\omega_0 + 2\Delta_{0,i}$, where $i$ is 1 or 2) as marked with magenta arrows, of which frequencies are located at local maximum slopes of the combined optical conductivity. In Fig. \ref{fig3}(C) we display the corresponding combined optical scattering rates obtained using the extended Drude model for the two $I^2\chi(\omega)$ functions. We note that we used the combined plasma frequency ($\simeq$ 1.8 eV) to obtain the scattering rate. We can see clearly the characteristic energy scales in the combined scattering rates as well. These characteristic energy scales can be seen in the combined reflectance spectra and the combined dielectric functions as well (not shown). We found from careful observation that these characteristic features and energy scales seem to appear in measured optical spectra of pnictides\cite{dai:2013a,min:2013}. So we tried to apply this approach to the real systems and obtain the electron-boson spectral density $I^2\chi(\omega)$ of the system. In the following section we described our applications to two pnictide systems.

\section{Applications to real systems}

\begin{figure}[!htbp]
  \vspace*{-0.3 cm}%
  \centerline{\includegraphics[width=4.0 in]{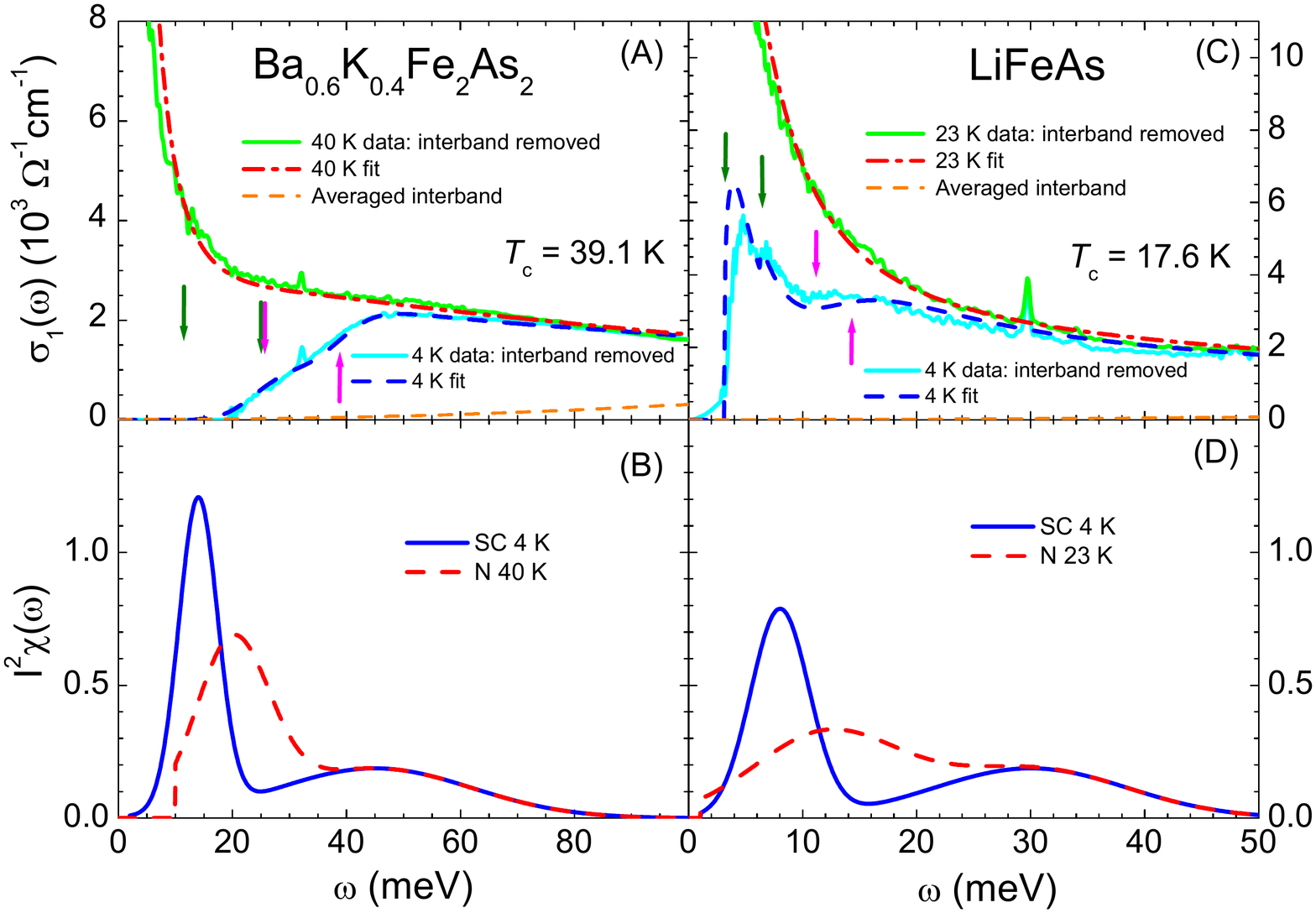}}%
  \vspace*{-0.2 cm}%
\caption{(A) The adjusted optical conductivity data and fits of Ba$_{0.6}$K$_{0.4}$Fe$_2$As$_2$ at two temperatures (40 K normal and 4 K superconducting states); the data are obtained from a published paper\cite{dai:2013a} (see in the text). The arrows indicate the characteristic energy scales: from left to right $2\Delta_{0,1}$, $2\Delta_{0,2}$, $2\Delta_{0,1}+\omega_{0,SC}$, and $2\Delta_{0,2}+\omega_{0,SC}$. We also display the averaged interband transition in a dashed orange line. (B) The obtained electron-boson spectral density functions from the fits at the two temperatures. (C) The adjusted optical conductivity data and fits of LiFeAs at two temperatures (23 K normal and 4 K superconducting states); the data are obtained from a published paper\cite{min:2013} (see in the text). The detailed fitting parameters are described in the text. The arrows indicate the characteristic energy scales: from left to right $2\Delta_{0,1}$, $2\Delta_{0,2}$, $2\Delta_{0,1}+\omega_{0,SC}$, and $2\Delta_{0,2}+\omega_{0,SC}$. We also display the averaged interband transition in a dashed orange line. (D) The obtained electron-boson spectral density functions from the fits at the two temperatures. We note that the scales of two horizontal axes are different.}
 \label{fig4}
\end{figure}

Here we applied our method to two multiband pnictide systems (Ba$_{0.6}$K$_{0.4}$Fe$_2$As$_2$ (BKFA) and LiFeAs) and obtained the electron-boson spectral density functions, $I^2\chi(\omega)).$ We digitized published reflectance spectra\cite{dai:2013a} of BKFA sample and obtained reflectance data of LiFeAs sample from one of the authors of a published paper\cite{min:2013}. And then we performed a Kramers-Kronig analysis independently to get the optical constants including the optical conductivity from the reflectance spectra. It has been known that the iron-pnictide materials including our two systems have low-energy interband transitions down to $\sim$20 meV energy levels\cite{min:2013,benfatto:2011,dai:2013,marsik:2013,lee:2015,dai:2015}. We need to remove the interband transitions before application of the method since we apply our method to the optical data up to 100 meV. However, the shape and intensity of the low-energy interband transition in the optical conductivity are not well-established yet. Therefore, we surveyed optical data published of four related pnictide systems (Co-, Ni-, and K-doped BaFe$_2$As$_2$, and LiFeAs)\cite{dai:2013,marsik:2013,lee:2015,dai:2015}. We separated the interband transition components and averaged them; the averaged interband transition is displayed in dashed orange lines in Fig. \ref{fig4}(A) and \ref{fig4}(C). Here we note that we do not use interband transitions of Min {\it et al.} since their Drude modes seem to be too narrow (confined below 1000 cm$^{-1}$) compared with results of other analyses which we use. We subtracted the averaged interband transition from the optical conductivity spectra of BKFA and LiFeAs to remove interband contributions. The adjusted optical conductivity data and fits of KBFA and LiFeAs are displayed in Fig. \ref{fig4}(A) and \ref{fig4}(C), respectively. We note that the interband transition is not very large compared to the charge-carrier contributions in the spectral ranges of interest.

We used a model $I^2\chi(\omega)$ which consists of two Gaussian modes (one is sharp and temperature-dependent and the other broad and temperature-independent). Here we used a broad Gaussian mode instead of the MMP mode since the broad Gaussian mode has one more parameter and is easier to describe a mode confined in a spectral range than the MMP mode. However, we expect that one can use a combination of a sharp Gaussian mode and an MMP mode to obtain similar results as ours. In the case of the combination of a Gaussian peak and an MMP mode one needs to impose a high-energy cutoff since the MMP mode spreads over a very wide spectral region. We note that even though we impose constraints on the shape of $I^2\chi(\omega)$ we still have some flexibility in the shape of resulting $I^2\chi(\omega)$ and can determine a most probable $I^2\chi(\omega)$ under the constraints through our fitting process. We have total 12 fitting parameters (6 for two Gaussian modes of the model $I^2\chi(\omega)$, 2 impurity scattering rates, 2 plasma frequencies, and 2 superconducting (SC) gaps for two transport channels) for SC state and 10 fitting parameters (here we do not need 2 SC gaps) for normal state.

We used known superconducting gaps of the two materials\cite{dai:2013a,min:2013}: two superconducting gaps of BKFA are 5.5 meV ($\Delta_{0,1}$) and 12.5 meV ($\Delta_{0,2}$)\cite{dai:2013a} and two SC gaps of LiFeAs are 1.6 meV and 3.15 meV\cite{min:2013}. We took the sharp mode frequencies from reported values by inelastic neutron scattering study of BKFA\cite{christianson:2008} and scanning tunneling\cite{chi:2012}, and optical spectroscopy studies\cite{hwang:2015}. The obtained $I^2\chi(\omega)$ are shown in Fig. \ref{fig4}(B) and \ref{fig4}(D): for superconducting case the sharp and broad Gaussian modes are located at, respectively, 14 meV (denoted as $\omega_{0,SC}$) and 45 meV for BKFA\cite{christianson:2008} and at, respectively, 8 meV (denoted as $\omega_{0,SC}$) and 30 meV for LiFeAs\cite{hwang:2015}. For normal state the sharp mode ($\omega_{0,SC}$) in $I^2\chi(\omega)$ for SC state should be broadened and moved to higher energies (20 meV ($\omega_{0,N}$) for BKFA and 12 meV for LiFeAs) to give less pronounced curvature in the calculated conductivity at normal state (refer to Fig. \ref{fig2}(B)). We needed a rather high low-energy cutoff (10 meV) in $I^2\chi(\omega)$ to capture the upturn at low energy region. If we do not have the cutoff the calculated upturn moves to higher energy and deviates significantly from the measured data. This cutoff seems to be dependent on material systems; LiFeAs shows lower cutoff energy (see Fig. \ref{fig4}(D)). The higher cutoff energy seems to give the large coupling constant ($\lambda$) and the higher logarithmically averaged frequency ($\omega_{ln}$). The definitions of these two new quantities are given in the last paragraph of this section. We do not know clearly physical origin of this low-energy cutoff at normal state yet.

While we did not need impurity scattering rates for both channels of BKFA we needed the impurity scattering rates for the two channels ($1/\tau_{imp,1} = 1/\tau_{imp,2} =$ 6.0 meV) of LiFeAs. Our impurity scattering rate of LiFeAs is consistent with a previous reported value of the same system\cite{hwang:2015} and indicates that the LiFeAs crystal may contain small amount of impurities. It is worth noting that two different approaches (one is with two Drude modes\cite{min:2013,dai:2013,marsik:2013,lee:2015,dai:2015,tu:2010,nakajima:2014} and the other ours with two extended Drude modes) may give different fitting parameters, particularly the impurity scattering rates. Our approach explicitly includes the inelastic scattering between charge carriers in the model calculations. The results of the approach with two Drude modes show that at low temperature the two Drude modes have very different scattering rates: one Drude channel is in the clean limit while the other is in the dirty limit\cite{wu:2010,dai:2015,tu:2010,nakajima:2014}. In our approach with two extended Drude modes we have non-zero scattering rate at zero frequency at finite temperatures and these values are proportional to the plasma frequency squared, and have additional scattering rate at finite frequencies. Therefore, one may not be able to compare the resulting parameters obtained from the two different approaches trivially. The fitting plasma frequencies are 10,500 cm$^{-1}$ ($\omega_{p,1}$) and 12,800 cm$^{-1}$ ($\omega_{p,2}$) for normal state and 9800 cm$^{-1}$ and 12,300 cm$^{-1}$ for corresponding SC state of BKFA, and 11,200 cm$^{-1}$ and 5800 cm$^{-1}$ for normal state and 10,500 cm$^{-1}$ and 5300 cm$^{-1}$ for SC state of LiFeAs. These plasma frequencies are not very different from reported Drude plasma frequencies obtained by two-Drude analyses of the both sample systems\cite{min:2013,dai:2013}. We note that the impurity scattering rates and the plasma frequencies cannot be determined independently; they seem to be inter-related to each other somehow in our method. The fits seem to match quite well with the measured conductivity spectra. The arrows show the positions of the characteristic energy scales as we have shown in Fig. \ref{fig3}(B): the dark green arrows indicate the SC gaps ($2\Delta_{0,1}$ and $2\Delta_{0,2}$) and the magenta arrows the sharp SC GP mode frequencies ($\omega_{0,SC}$) combined with the SC gaps ($2\Delta_{0,1} + \omega_{0,SC}$ and $2\Delta_{0,2} + \omega_{0,SC}$).

\begin{figure}[!htbp]
  \vspace*{-0.3 cm}%
  \centerline{\includegraphics[width=4.0 in]{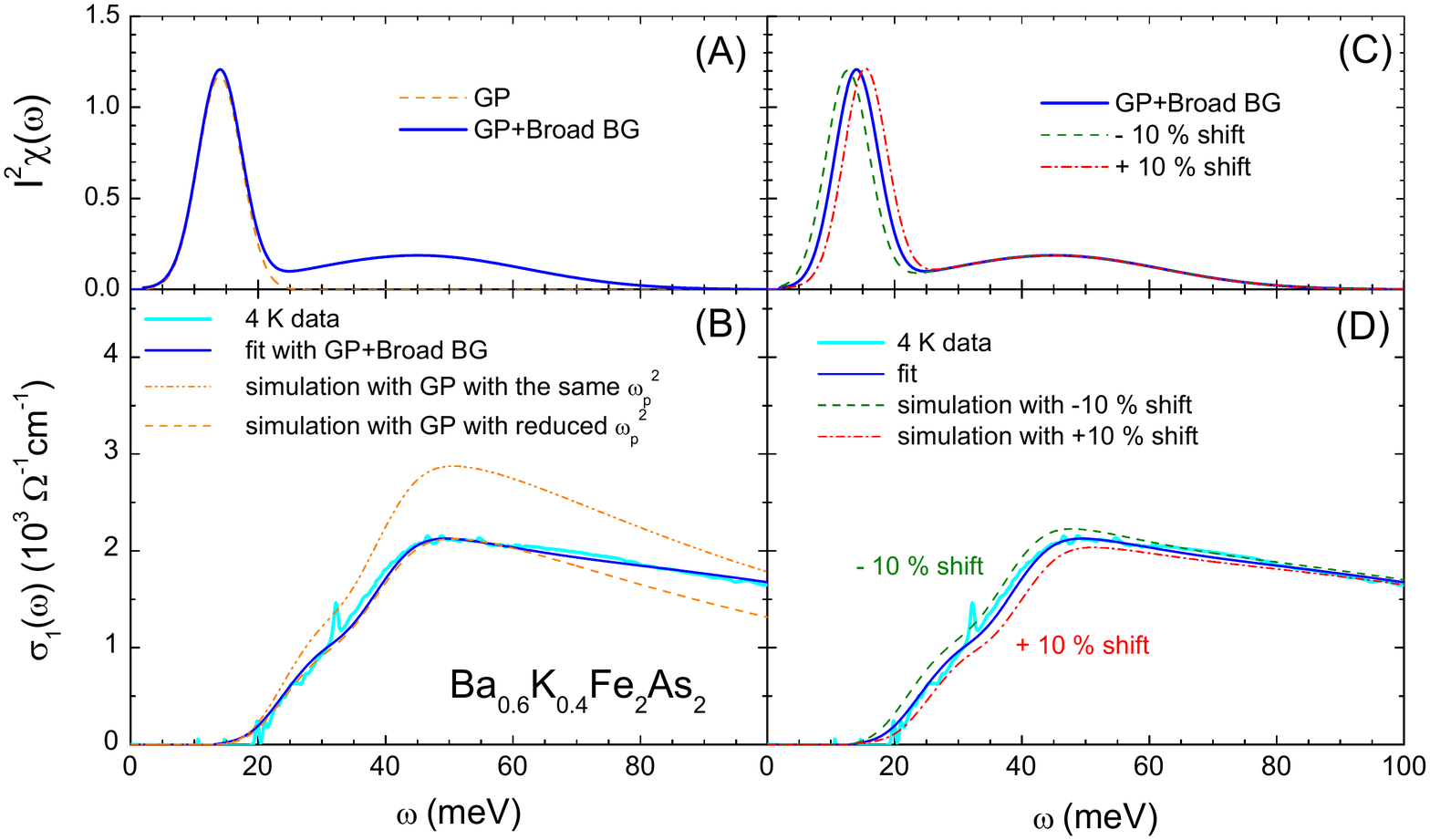}}%
  \vspace*{-0.7 cm}%
\caption{(A) Two $I^2\chi(\omega)$: one consists of a sharp Gaussian peak (denoted as GP) and the other the GP mode plus a broad Gaussian background (BG) (denoted as GP+Broad BG). (B) The adjusted optical conductivity of Ba$_{0.6}$K$_{0.4}$Fe$_2$As$_2$ (BKFA) sample at SC state (4 K) and comparison of the resulting combined optical conductivity spectra obtained using the two different $I^2\chi(\omega)$. (C) Obtained $I^2\chi(\omega)$ at 4 K and two others shifted the sharp GP by $\pm$10 \% (or $\pm$ 1.4 meV). (D) The adjusted optical conductivity of BKFA sample at SC state (4 K) and comparison of the resulting combined optical conductivity spectra obtained using the $I^2\chi(\omega)$ shown in panel (C).}
 \label{fig4add}
\end{figure}

We have two important remarks on our fitting. First, we note that could not fit the adjusted data well with a single Gaussian mode. Here we consider the adjusted data of BKFA sample at superconducting state (4 K). In Fig \ref{fig4add}(A) we show two $I^2\chi(\omega)$: one is the sharp Gaussian peak (denoted as GP) and the other the GP plus a broad Gaussian background (BG) (denoted as GP+Broad BG). In Fig \ref{fig4add}(B) the double dot-dashed orange line is obtained using the GP $I^2\chi(\omega)$ alone with the same plasma frequencies ($\omega_{p,1} =$ 9800 cm$^{-1}$ and $\omega_{p,1} =$ 12,300 cm$^{-1}$) of the two channels as the ones which we used to get the solid blue line, which is our fit to the data. The dashed orange line is obtained using the same GP but smaller plasma frequencies ($\omega_{p,1} =$ 8500 cm$^{-1}$ and $\omega_{p,1} =$ 10,500 cm$^{-1}$) to match the combined conductivity with the data in low energy region. But in this case we still cannot fit the high energy region above 60 meV. So to fit the whole spectral range of interest we have to add the broad Gaussian BG mode in the model $I^2\chi(\omega)$ as we can see that the fit with GP+Broad BG (blue solid line) shows good match with the data (cyan solid line). Interestingly, the additional Gaussian BG mode improves slightly the fit between 20 and 40 meV as well. Secondly, we estimate the accuracy of the sharp Gaussian peak position in the mode GP+Broad BG $I^2\chi(\omega)$. Fig \ref{fig4add}(C-D) show how sensitive (or robust) the main sharp Gaussian peak (GP) position is to the fit. Here again we consider the adjusted data of BKFA at superconducting state. We observe that when we shift the sharp GP mode located at 14 meV by $\pm$10\% (or $\pm$ 1.4 meV) the resulting combined optical conductivity is significantly affected by the shifts; -10\% and +10\% peak position shifts result in significant ($\pm \sim 7\%$ at 1000 $\Omega^{-1}$cm$^{-1}$ level) shifts of the conductivity edge to lower (for -10\% shift) and higher (for +10\% shift) energies. We estimate that the accuracy of the Gaussian peak position in our fitting is within $\pm \sim$5\% (or $\pm \sim$0.7 meV) or better.

\begin{figure}[]
  \vspace*{-0.7 cm}%
  \centerline{\includegraphics[width=3.5 in]{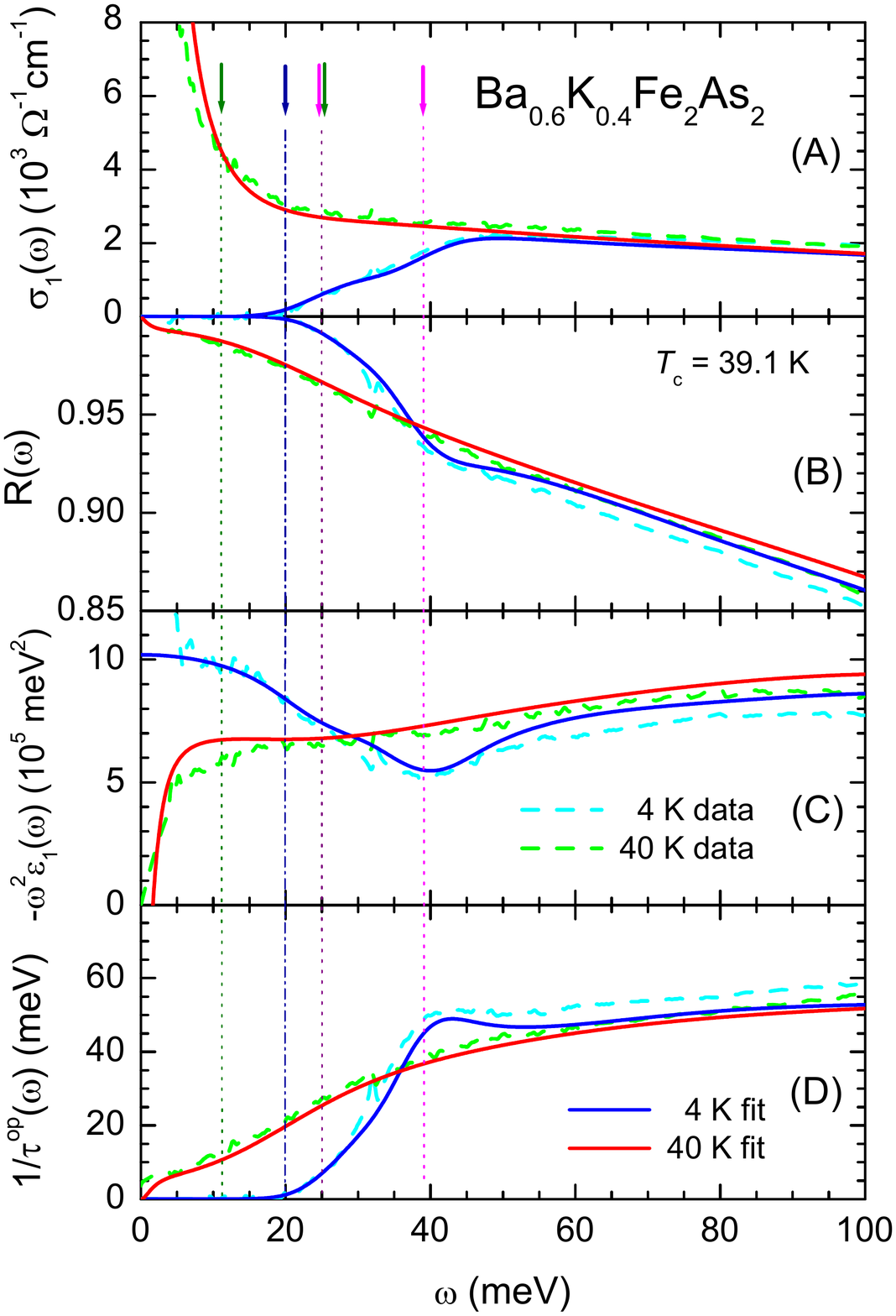}}%
  \vspace*{-0.7 cm}%
\caption{The data and fits of the BKFA sample at normal (40 K) and superconducting (4 K) states for four optical quantities: (A) the optical conductivity ($\sigma_1(\omega)$), (B) reflectance spectra ($R(\omega)$), (C) the dielectric function ($-\omega^2\epsilon_1(\omega)$), and (D) the optical scattering rates ($1/\tau^{op}(\omega)$). We note that the all data here are not adjusted. The arrows and vertical lines indicate the characteristic energy scales: from left to right $2\Delta_{0,1}$, $\omega_{0,N}$, $2\Delta_{0,2}$, $2\Delta_{0,1}+\omega_{0,SC}$, and $2\Delta_{0,2}+\omega_{0,SC}$ are 11, 14, 25, 25, and 39 meV, respectively.}
 \label{fig5}
\end{figure}

\begin{figure}[]
  \vspace*{-0.7 cm}%
  \centerline{\includegraphics[width=3.5 in]{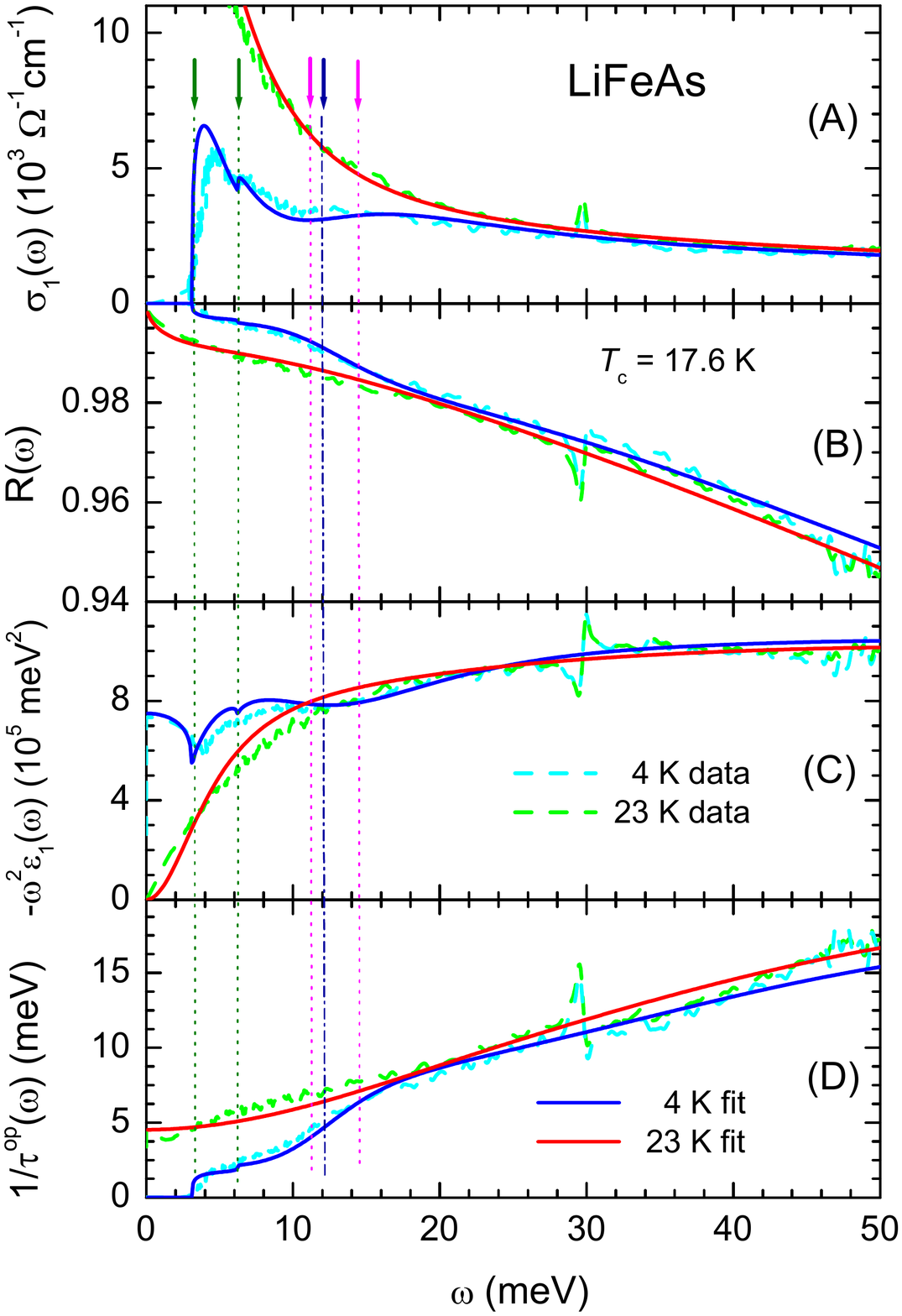}}%
  \vspace*{-0.7 cm}%
\caption{The data and fits of the LiFeAs sample at normal (23 K) and superconducting (4 K) states for four optical quantities: (A) the optical conductivity ($\sigma_1(\omega)$), (B) reflectance spectra ($R(\omega)$), (C) the dielectric function ($-\omega^2\epsilon_1(\omega)$), and (D) the optical scattering rates ($1/\tau^{op}(\omega)$). We note that the all data here are not adjusted. The arrows and vertical lines indicate the characteristic energy scales: from left to right $2\Delta_{0,1}$, $2\Delta_{0,2}$, $\omega_{0,N}$, $2\Delta_{0,1}+\omega_{0,SC}$, and $2\Delta_{0,2}+\omega_{0,SC}$ are 3.2, 6.3, 8.0, 11.2, and 14.3 meV, respectively.}
 \label{fig6}
\end{figure}

To see validity of our fits further we calculated and compared other combined optical quantities using the fits to the adjusted optical conductivity data shown in Fig. \ref{fig4}(A) and \ref{fig4}(C). Other quantities are reflectance ($R(\omega)$), the dielectric function ($-\omega^2\epsilon_1(\omega)$), and the optical scattering rate ($1/\tau^{op}(\omega)$). In Fig. \ref{fig5}(A)-(D) we display data and fits of the four optical quantities including the combined optical conductivity of the BKFA sample at normal and superconducting states. We note that the all data here are raw spectra (not adjusted). We described the formalism which we used to get those quantities in the subsection of "The combined optical quantities from the fitting". For the normal case at 40 K of BKFA sample with no impurity scattering we had to take into account the spectral weight which appears at zero frequency due to no impurity scattering; since we dealt optical data at normal state we needed to subtract the contribution [$2\:SW_{@0}/(\pi \omega)$] of the spectral weight at $\omega =$ 0 from the imaginary part of the optical conductivity, where $SW_{@0}$ stands for the spectral weight at $\omega =$ 0. We used the spectral weight as 6.88$\times$10$^6$ $\Omega^{-1}$cm$^{-2}$; the corresponding plasma frequency is 0.92 eV or 7418 cm$^{-1}$, which is smaller than but comparable to the superfluid plasma frequency. We note that the superfluid plasma frequency obtained from the measured data is 1.01 eV or 8146 cm$^{-1}$. We used the background dielectric constant ($\epsilon_H$ of Eq. (\ref{eq7})) as 53 which is consistent with a reported value of the same system\cite{charnukha:2011}. Overall agreements between data and fits for all four quantities are good even though we do not include the low-energy interband transitions in the fits. This means that the two-independent complete optical data sets (for example, $\sigma_1(\omega)$ and $\epsilon_1(\omega)$) are fitted simultaneously with the same $I^2\chi(\omega)$ and the fitting parameters, which may indicate that our fittings are reliable. We can see the characteristic energy scales which are marked with arrows and vertical lines. We note that here the plasma frequency, which we used for calculating the optical scattering rate, can be any value as long as we used the same value for both data and fits. The plasma frequencies are 1.2 eV for BKFA and 1.0 eV for LiFeAs, respectively. Since we have zero impurity scattering for BKFA sample the SC gap features do not appear clearly in all four quantities. In Fig. \ref{fig6}(A)-(D) we display data and fits of the four optical quantities of LiFeAs sample for normal (23 K) and superconducting (4 K) cases. We also note that the all data here are raw spectra (not adjusted). The superfluid plasma frequency obtained from the measured data is 0.87 eV or 7016 cm$^{-1}$, which is similar to a reported value of the same system\cite{min:2013}. Since we have a finite impurity scattering rate (6 meV) the SC gap features appear clearly in all four optical quantities. We used the background dielectric constant ($\epsilon_H$) as 103, which is larger than that of BKFA; this might be related to lower energy scales of LiFeAs compared with those of BKFA. Overall agreements between data and fits are quite good.

The coupling constant ($\lambda \equiv 2\int_0^{\omega_c} I^2\chi(\Omega)/\Omega\: d\Omega$, where $\omega_c$ is the cutoff frequency) can be obtained from the obtained $I^2\chi(\omega)$. The coupling constants are 1.98 for superconducting and 1.83 for normal states of BKFA and corresponding values of LiFeAs are 1.72 (SC) and 1.37 (normal). The coupling constant at SC state is larger than that at normal state. We crudely estimated the maximum SC transition temperatures ($T_c^{max}$) using a generalized McMillan equation ($k_B T_c^{max} \cong 1.13\: \hbar \omega_{ln} \exp[-(\lambda+1)/\lambda]$, where $\omega_{ln}$ ($\equiv \exp[(2/\lambda)\int_0^{\omega_c}\ln{\Omega}\: I^2\chi(\Omega)/\Omega\: d\Omega$]) is the logarithmically averaged frequency). The maximum $T_c$ estimated are 44.1 K and 52.2 K for BKFA at 4 K and 40 K, respectively, and 21.9 K and 21.2 K for LiFeAs at 4 K and 23 K, respectively. The maximum SC transition temperatures are higher than the measured $T_c$ of the two samples: $T_c =$ 39.1 K for BKFA and 17.6 K for LiFeAs. The electron-boson spectrum obtained may be enough to give the superconductivity in these materials. We note that the McMillan formula was derived for a single gap system\cite{mcmillan:1968}. When we eliminate the broad Gaussian background from the obtained $I^2\chi(\omega)$ and recalculate the maximum $T_c$'s we have lower maximum $T_c$'s than those obtained with including the broad Gaussian background as 31.4 K and 34.6 K at 4 K and 40 K for BKFA sample, respectively, and the corresponding maximum $T_c$'s as 15.3 k and 13.1 K at 4 K and 23 K for LiFeAs sample, respectively. For the both samples the new maximum $T_c$'s are lower than the actual $T_c$'s. Therefore, the broad Gaussian background is necessary to give high enough maximum $T_c$'s; the electron-boson spectral density necessarily consists of a sharp Gaussian peak and an additional broad Gaussian background. The mode energy of the sharp Gaussian peak of $I^2\chi(\omega)$ at SC state can be scaled to $T_c$ ($\omega_{0,SC} = $ 4.1$k_B T_c$ for BKFA and 5.2$k_B T_c$ for LiFeAs) as the magnetic resonance mode in cuprates ($E_{res} = $ 5.4$k_B T_c$), where $E_{res}$ is the magnetic resonant mode energy\cite{he:2001}. The sharp Gaussian mode in $I^2\chi(\omega)$ moves to higher energy and becomes broadened as the sample temperature ($T$) increases. The $T_c$- and $T$-dependent properties of the obtained $I^2\chi(\omega)$ are similar to those of cuprates\cite{carbotte:2011}.

\section{Conclusions}

We introduced an approximate optical method which could be used to obtain the electron-boson spectral density function from the optical conductivity data of correlated multiband systems. We found that the combined conductivity of two transport channels obtained using this method clearly shows characteristic features of both channels, particularly different superconducting gaps and sharp peaks in the electron-boson spectral density function ($I^2\chi(\omega)$). We also observed that similar characteristic features appear in measured optical spectra of multiband pnictide systems. So we applied this method to measured optical conductivity spectra of two iron-pnictide systems, BKFA and LiFeAs and obtained the electron-boson spectral density functions, $I^2\chi(\omega)$ of the two multiband systems. The obtained $I^2\chi(\omega)$ showed a common shape (a sharp peak and a broad background) and similar temperature and superconducting transition temperature ($T_c$) dependencies as $I^2\chi(\omega)$ of cuprates does. As in cuprates many believe that in iron-pnictides the mediated bosons are associated with the antiferromagnetic spin-fluctuations since the superconducting phase is located nearby the antiferromagnetic one in the phase diagram. We expect that our method may open a new avenue for analyzing optical data of correlated multiband systems and help to reveal the origin of strong correlations in the systems.

%
% Acknowledgments
%

\ack
The author acknowledges financial support from the National Research Foundation of Korea (NRFK Grant No. 2013R1A2A2A01067629).

%
% bibliography
%
\section*{References}
\bibliographystyle{unsrt}
\bibliography{bib}% Produces the bibliography via BibTeX.

\end{document}